\documentclass[manuscript]{aastex}

\usepackage{graphicx}
\usepackage{epstopdf}

\shorttitle{Tholin Density}
\shortauthors{H\"orst and Tolbert}

\begin{document}


\title{\emph{In Situ} Measurements of the Size and Density of Titan Aerosol Analogues}

\author{S.M. H\"orst\altaffilmark{1}, 
M.A Tolbert\altaffilmark{1,2}}
\altaffiltext{1}{Cooperative Institute for Research in Environmental Sciences, University of Colorado, Boulder, CO, USA}
\altaffiltext{2}{Department of Chemistry and Biochemistry, University of Colorado, Boulder, CO, USA}
\email{sarah.horst@colorado.edu}


%\altaffiltext{1}{Joint Appointment, Lunar and Planetary Laboratory, The University of Arizona,1629 E. University Blvd., Tucson, AZ 85721, USA.}


\begin{abstract}
The organic haze produced from complex CH$_{4}$/N$_{2}$ chemistry in the atmosphere of Titan plays an important role in processes that occur in the atmosphere and on its surface. The haze particles act as condensation nuclei and are therefore involved in Titan's methane hydrological cycle. They also may behave like sediment on Titan's surface and participate in both fluvial and aeolian processes. Models that seek to understand these processes require information about the physical properties of the particles including their size and density. Although measurements obtained by Cassini-Huygens have placed constraints on the size of the haze particles, their densities remain unknown. We have conducted a series of Titan atmosphere simulation experiments and measured the size, number density, and particle density of Titan aerosol analogues, or tholins, for CH$_{4}$ concentrations from 0.01\% to 10\% using two different energy sources, spark discharge and UV. We find that the densities currently in use by many Titan models are higher than the measured densities of our tholins. 
\end{abstract}





\section{Introduction}

The atmosphere of Titan, Saturn's largest moon, possesses a characteristic organic haze that results from photochemistry initiated by the destruction of N$_{2}$ and CH$_{4}$ in Titan's upper atmosphere. These haze particles can serve as condensation nuclei in the atmosphere and eventually settle to the surface where they may play a key role in both fluvial and aeolian processes. The extensive dune fields on Titan's surface are most likely composed of organic materials that originate from the atmosphere \citep{Soderblom:2007} and organic materials may also act as sediment transported by Titan's extensive stream and river systems \citep{Burr:2006}.

Measurements acquired by the Cassini-Huygens Spacecraft, particularly the Descent Imager/Spectral Radiometer (DISR) carried by Huygens to the surface, have improved our constraints on the physical properties of the haze particles, specifically their size and number density \citep{Tomasko:2005, Tomasko:2008}. However, the particle density remains unknown. Particle density is an important parameter in models of both the atmosphere and the surface including aerosol/cloud microphysics models and fluvial and aeolian transport models. Most microphysics models use a haze primary particle density of 1 g/cm$^{3}$ without experimental or observational constraints \citep{Toon:1992, Lavvas:2011}, while surface models often use higher values for organic particles up to 1.5 g/cm$^{3}$ \citep{Burr:2006} which came from material density measurements of poly-HCN \citep{Khare:1994} and bitumen \citep{Moroz:1998}.

Although Titan atmosphere simulation experiments have produced Titan aerosol analogues, so-called ``tholins'', for nearly 50 years, experimental measurements of the densities of the particles produced in these experiments are lacking. Previous density measurements have been reported of tholins produced using FUV photons to irradiate 0.1\% CH$_{4}$ in N$_{2}$ \citep{Trainer:2006} and cold plasma to irradiate 10\% CH$_{4}$ in N$_{2}$ \citep{Imanaka:2012}. Although the current abundance of CH$_{4}$ is $\sim$2\% \citep{Flasar:2005,Cui:2009}, the irreversible destruction of CH$_{4}$ in Titan's atmosphere almost certainly leads to a variation of the CH$_{4}$ abundance over time. Photons and energetic electrons both contribute to the formation of haze in Titan's atmosphere. While it is impossible to perfectly replicate the energy environment in the laboratory, here we explore the effect of energy source on the physical properties of laboratory produced analogues. We present here measurements of the size, number density, and particle density of tholins produced using a spark discharge or UV photons to irradiate CH$_{4}$ concentrations (denoted [CH$_{4}$])  from 0.01\% to 10\%. These results may also be of interest for understanding exoplanet atmospheric hazes. 

%burr 2012 gas relatively unconstrained
	
\section{Materials and Experimental Methods}
\subsection{Haze Production Setup}

\begin{figure}
\plotone{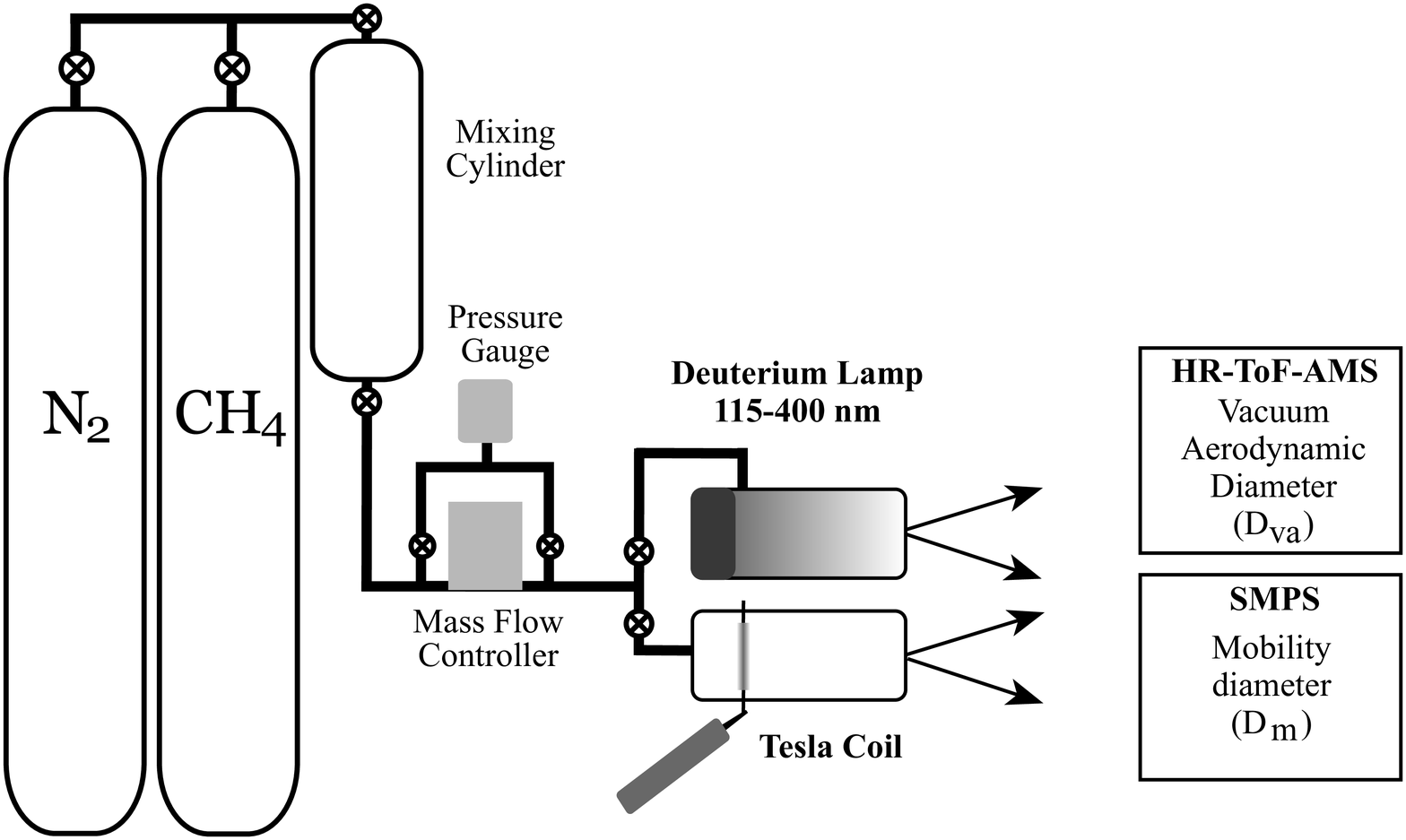}
\caption{Schematic of the experimental setup. N$_{2}$ and CH$_{4}$ mix overnight in the mixing cylinder. Gases flow through the UV or spark reaction cell where they are exposed to FUV photons from a deuterium lamp or the electric discharge produced by a tesla coil initiating chemical processes that lead to the formation of new gas phase products and particles. The particles are analyzed using either a high resolution time-of-flight aerosol mass spectrometer (HR-ToF-AMS) to measure their vacuum aerodynamic diameter or a scanning mobility particle sizer (SMPS) to measure their size distribution. Each reaction cell has two outlets. \label{fig:experiment}}
\end{figure}

We introduced CH$_{4}$ (99.99\% Airgas) in volume mixing ratios ranging from 0.01\% to 10\% (see Table 1) into a stainless steel mixing chamber, then filled the mixing chamber to 600 PSI with N$_{2}$ (99.999\% Airgas). The gases mix for a minimum of 8 hours. The reactant gases continuously flowed through a glass reaction cell at 100 standard cubic centimeters per minute (sccm), a rate determined by instrument requirements, using a mass flow controller (Mykrolis FC-2900). The pressure in the reaction cell was maintained between 620 and 640 Torr at room temperature. We expose the reactant gases to one of two energy sources, spark discharge or UV. The setup is identical for both types of experiment until the gases reach the reaction cells. The spark reaction cell is connected to a tesla coil (Electro Technic Products). The UV reaction cell is connected to a deuterium continuum lamp (Hamamatsu L1835, MgF$_{2}$ window). The flow continues out of the cell and into either a high resolution time-of-flight aerosol mass spectrometer (HR-ToF-AMS) or a scanning mobility particle sizer (SMPS). A schematic of the experimental setup is shown in Figure \ref{fig:experiment}. Previous spark and UV experiments were performed using a similar setup by \citet{Trainer:2004, Trainer:2004b} and by \citet{Trainer:2006,Trainer:2012,Trainer:2013}, respectively. 

The energy from the tesla coil or deuterium lamp initiates the chemistry that results in the formation of aerosol. We use the electrical discharge because it is known to dissociate N$_{2}$. Although the energy density is higher than the energy available to initiate chemistry in Titan's atmosphere, it is used here as an analogue for the relatively energetic environment of Titan's upper atmosphere. The tesla coil operates at a range of voltages and was set to minimize the energy density while still producing sufficient aerosol using 2\% CH$_{4}$ for our analytical techniques. The same setting was used for all experiments.

Although energetic electrons and particles contribute to Titan's atmospheric chemistry, photons are the dominant source of energy for ionization and dissociation \citep{Lavvas:2011b}. We are therefore also interested in exploring these processes. The FUV photons produced by the deuterium lamp (115-400 nm) are not sufficiently energetic to directly dissociate N$_{2}$; however, \citet{Trainer:2012} demonstrated that nitrogen is participating in the chemistry in our reaction cell, most likely through processes involving the products of CH$_{4}$ dissociation. 

One advantage of our \emph{in situ} techniques is that all measurements are obtained without exposing the particles to Earth's atmosphere and without collecting the samples on a surface, which could affect the observed particle sizes and densities. However, production rates must be high enough to produce sufficient aerosol for real time analysis; for this reason we typically run our experiments at 620-640 Torr (atmospheric pressure in Boulder, Colorado, altitude $\sim$1600 m). These pressures are higher than in the regions of Titan's atmosphere where the chemistry leading to aerosol formation begins. For that reason, work is ongoing in our laboratory to investigate any pressure dependence in our results and to ascertain the range of pressures for which experiments can be run based on our experimental and analytical constraints. For the purposes of this work, we are interested in comparing differences resulting only from choice of energy source and gas mixing ratio at our standard experimental pressure. 

\subsection{High Resolution Time-of-Flight Mass Spectrometry (HR-ToF-AMS)}

We use a high-resolution time-of-flight aerosol mass spectrometer (HR-ToF-AMS, or AMS, Aerodyne Research) operating in particle time-of-flight (PToF) mode to measure mass spectra as a function of flight time of the particles \citep{decarlo:2006}. Briefly, the particles exit the reaction chamber, flow through a critical orifice (120 $\mu$m) and are focused by an aerodynamic lens transmitting particles with aerodynamic diameters ($D_{va}$) from $\sim$20 nm to $\sim$1 $\mu$m into the particle time-of-flight region. The particles are flash vaporized at $\sim$600$^{\circ}$C. The resulting molecules are ionized via 70 eV electron ionization. Ions are analyzed by a high-resolution time-of-flight mass spectrometer (H-TOF Platform, Tofwerk). In PToF mode, the size dependent velocities of the particles in the particle time-of-flight region, obtained by chopping the particle beam to allow for arrival time measurements, are used to determine the particle vacuum aerodynamic diameter ($D_{va}$) \citep{Jimenez:2003a, Jimenez:2003b, decarlo:2006}. $D_{va}$ is defined as the diameter of a unit density sphere that reaches the same terminal velocity in the AMS as the measured particle. In this work, we use the PToF mode to measure $D_{va}$, which in combination with $D_{m}$ determined by SMPS measurements allows for particle density calculations. The AMS data were analyzed using AMS analysis software programs SQUIRREL and PIKA \citep{decarlo:2006, Aiken:2007, Aiken:2008}.

\subsection{Scanning Mobility Particle Sizer (SMPS)}

We use a scanning mobility particle sizer (SMPS) to measure the particle size distribution. The SMPS consists of a electrostatic classifier (TSI 3080), a differential mobility analyzer (DMA, TSI 3081), and a condensation particle counter (CPC, TSI 3775). The SMPS requires a higher flow rate than the AMS; we add a flow of N$_{2}$ after the reaction chamber to bring the total flow rate to 260 sccm. The polydisperse aerosol enters the DMA, which applies an electric field to the flow of particles and size selects them based on their electrical mobility against the drag force provided by the sheath flow. The size-selected particles enter the CPC, which measures the number of particles by light scattering. The SMPS provides the number of particles as a function of their mobility diameter ($D_{m}$). We used sheath flows of 3 L/min or 10 L/min ($D_{m}$ ranges of 14.5-673 nm or 7.4-289 nm, respectively). We corrected for the dilution caused by the additional flow of N$_{2}$. Due to experimental flow rate constraints, the SMPS and AMS measurements were obtained consecutively, rather than simultaneously, by switching between outlets to the SMPS and AMS once particle production stabilized.

\section{Particle Mass Loading, Size, and Density \label{sect:den}}

The AMS measurements of aerosol mass loading as a function of initial [CH$_{4}$] are shown in Figure \ref{fig:allfour}A for both the spark and UV experiments. Shown also is the total volume loading as measured by the SMPS. The UV results are consistent with \citet{Trainer:2006}. The energy source affects the [CH$_{4}$] at which the peak in aerosol production occurs ($\sim$2\% CH$_{4}$ for the spark discharge compared to 0.01\% CH$_{4}$-0.1\% CH$_{4}$ for the UV). Other plasma experiments result in similar production curves; \citet{Sciamma:2010} saw peak aerosol production at 4\% and 6\% CH$_{4}$ for experiments run at 0.9 and 1.7 mbar, respectively, using a cold plasma discharge. They suggested that the initial increase in aerosol production as CH$_{4}$ increases results from aerosol formation being CH$_{4}$ limited at low abundances. However, at higher abundances an increase in production of H$_{2}$ and H with increasing CH$_{4}$ abundance may limit aerosol formation. In particular, they suggest based on N$_{2}$-H$_{2}$ experiments \citep{Loureiro:1993} that the presence of H$_{2}$ can decrease the N$_{2}$ dissociation efficiency and that nitrogen may play a key role in aerosol formation in these experiments.  

The trend observed in the UV experiments is almost certainly related to optical depth \citep{Trainer:2006}. At [CH$_{4}$] less than ~0.02\%, the entire reaction cell is optically thin at Lyman-$\alpha$. Therefore the decrease in production rate observed as a function of increasing [CH$_{4}$] shown in Figure \ref{fig:allfour}A is most likely due to the increase in optical depth. 

The measurements from the SMPS show that the trends observed in loading result both from a change in the number of particles produced (Figure \ref{fig:allfour}B) and a change in particle size (Figure \ref{fig:allfour}C). This indicates that the initial [CH$_{4}$] affects both particle formation and particle growth. For all gas mixtures investigated, the UV experiments produced more particles than the spark experiments. For the 2\% CH$_{4}$ experiments, the resulting aerosol diameters based on number density ($D_{m,N}$) were 29.5 $\pm$ 2.2 nm for the UV and 42.6 $\pm$ 1.9 nm for the spark. DISR measurements indicate that Titan's aerosols are likely fractal aggregates of approximately 3000 primary particles with a primary particle diameter of $\sim$100 nm \citep{Tomasko:2008}. Our aerosols are a little less than half the diameter of the primary particles in Titan's troposphere. 

\begin{figure}
\plotone{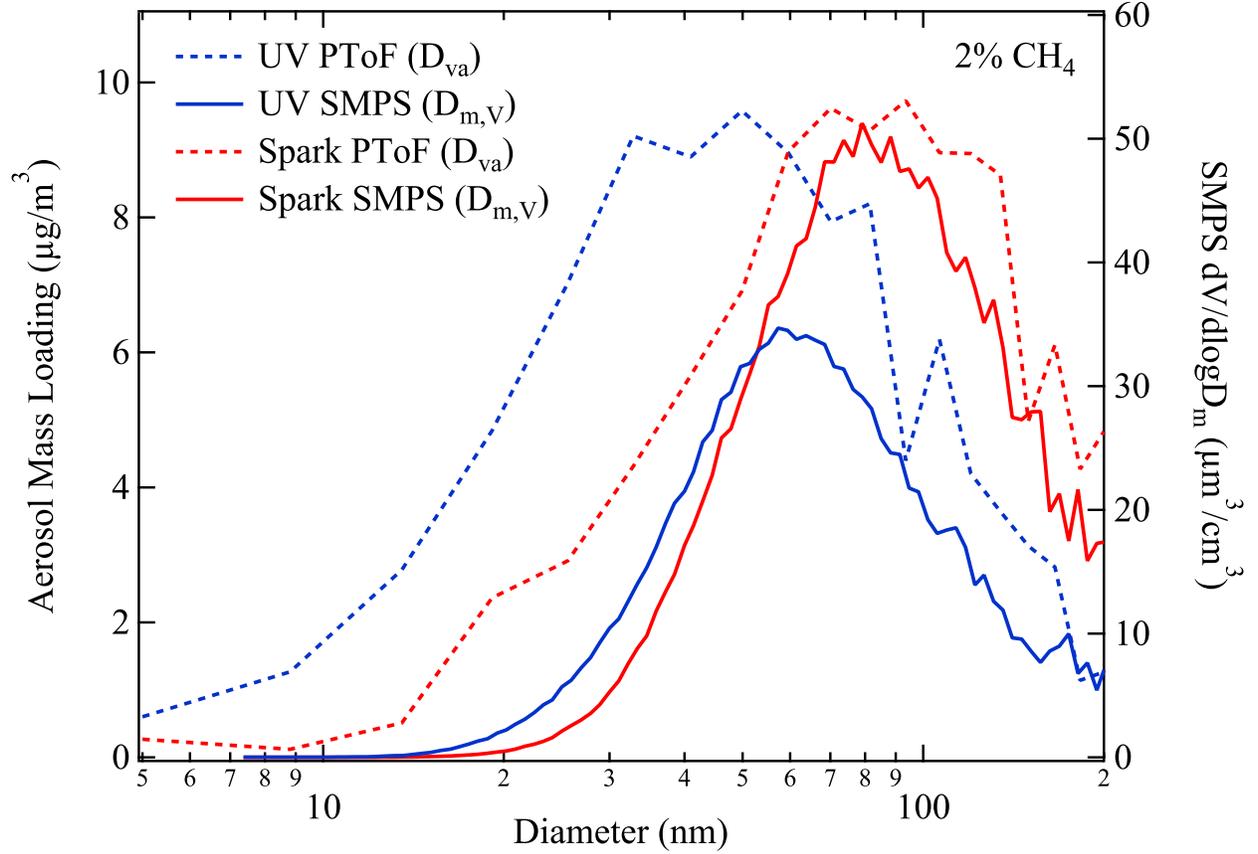}
\caption{AMS aerosol mass loading as a function of vacuum aerodynamic diameter ($D_{va}$) (left axis, dashed) and SMPS aerosol volume loading as a function of mobility diameter ($D_{m}$) (right axis, solid). If particles have $\rho_{eff}$ = 1 g/cm$^{3}$, the PToF and SMPS size distributions peak at the same diameter; peaks at different diameters indicate $\rho_{eff} \ne 1$ g/cm$^{3}$. \label{fig:ptof_smps}}
\end{figure}

The combination of size measurements from the AMS and SMPS provide the opportunity to constrain the particle density. The AMS PToF measurements yield the mass distribution as a function of vacuum aerodynamic diameter ($D_{va}$).  $D_{va}$ is affected both by the particle shape and density. The SMPS provides mobility diameter ($D_{m}$), which is also related to particle shape. Figure \ref{fig:ptof_smps} compares the PToF and SMPS distributions for aerosols made from 2\% CH$_{4}$ using UV or spark energy sources. Previous works have shown that the two diameters are related by effective particle density $\rho_{eff}$,
\begin{equation} \rho_{eff}=\rho_{0}\frac{D_{va}}{D_{m}}=\rho_{m}S \end{equation}
where $\rho_{0}$ is the unit density (1 g/cm$^{3}$), $\rho_{m}$ is the material density of the particle, and $S$ is the shape factor \citep{Decarlo:2004, Jimenez:2003a, Jimenez:2003b}. $S$ is a function of particle shape and fraction of internal voids. Therefore $\rho_{eff}$ is proportional to the material density, but depends on the shape and porosity. A spherical particle with no internal voids has an $S$ value of one \citep{Jayne:2000, Decarlo:2004}. Internal voids and/or irregular shape both decrease the value of $S$. We can also compare $\rho_{eff}$ to directly particle density $\rho_{p}$
\begin{equation} \rho_{eff}=\rho_{p}S' \end{equation}
where $S'$ is a modified shape factor that depends only on the shape of the particle and equals one for spherical particles. 
Therefore, $\rho_{eff}\le\rho_{p}\le\rho_{m}$ where the equality between $\rho_{p}$ and $\rho_{m}$ holds if there are no internal voids and the equality between $\rho_{eff}$ and $\rho_{p}$ holds if the particles are spherical. Table 2 summarizes the relationships between these densities. While the total mass measured by the AMS and total volume measured by the SMPS (Fig\ref{fig:allfour}A) can be used to calculate density, this requires extremely accurate mass calibration of the AMS. The method used here is more reliable \citep{decarlo:2006}.

For use in this calculation, $D_{m}$ is the average mobility diameter calculated using particle volume distribution (D$_{m,V}$), not the number density distribution (D$_{m,N}$), which was discussed above \citep{decarlo:2006}. Previous atomic force microscope (AFM) and transmission electron microscope measurements (TEM) of Titan tholins produced in our laboratory demonstrated that the particles are spherical \citep{Trainer:2006, Hasenkopf:2011}. Assuming that the particles are spherical, and contain no internal voids, the material densities produced in these experiments range from $\sim$0.5-1.1 g/cm$^{3}$ as shown in Figure \ref{fig:allfour}D and listed in Table 1. Two previous measurements of tholin densities have been reported. \citet{Trainer:2006} found a density of $\sim$0.8 g/cm$^{3}$ for tholins produced from UV irradiation of 0.1\% CH$_{4}$ using a similar experimental setup, which is consistent with the density reported here. \citet{Imanaka:2012} found densities of $\sim$1.3-1.4 g/cm$^{3}$ for tholins produced from a cold plasma discharge using 10\% CH$_{4}$ at pressures of 1.6 and 23 mbar. While our 2\% CH$_{4}$ spark particle density ($\sim$1.1 g/cm$^{3}$) approaches that range, our 10\% CH$_{4}$ particles have a much lower density ($\sim$0.4 g/cm$^{3}$). However, our experimental pressure is much higher than that of \citet{Imanaka:2012}, which may account for the difference. Additionally, their measurements used tholin deposited on a substrate before the density measurements were performed; the deposition process may increase the density if the particles are porous. If that is the case, the density measurements of \citet{Imanaka:2012} can be considered measurements of the material density ($\rho_{m}$) rather than the particle density ($\rho_{p}$), which increases the possibility that our results are consistent with theirs. Our density measurements are comparable to the densities of some suggested building blocks of tholins, which can range from 0.7 g/cm$^{3}$ (HCN) to 1 g/cm$^{3}$ (pyridine, C$_{5}$H$_{5}$N) at the temperature and pressure conditions of our lab \citep{Haynes:2012}. However, the density measurements do not place any new constraints on the chemical composition of our particles.

Both the spark and UV tholins demonstrate a variation in effective density as a function of initial [CH$_{4}$], and peak in approximately the same region-- 1-2\% CH$_{4}$. Surprisingly, with the exception of the 2\% CH$_{4}$ tholins, the energy source has little effect on the density. The spark tholin density trend is similar to the trends observed in diameter and number density, while the UV density trend is only similar to the diameter and number density trends for [CH$_{4}$] $\ge$ 1\% CH$_{4}$. Previous measurements using this experimental apparatus have shown that the composition of the UV tholins is independent of [CH$_{4}$] \citep{Trainer:2006}, indicating that $\rho_{m}$ is constant with [CH$_{4}$]. If we assume that all of the particles are spherical, based on the previous AFM and TEM measurements obtained in our laboratory, then the variation in $\rho_{eff}$ observed would result entirely from porosity variation; the 1\% CH$_{4}$ tholins would be the least porous of the UV produced particles. The porosity behavior may result from the availability of building blocks; at low [CH$_{4}$] particle chemistry is CH$_{4}$ limited and at high [CH$_{4}$] it is photon limited. The situation is more complex for the spark tholins where previous work in our lab has shown that the composition does depend on [CH$_{4}$] \citep{Trainer:2004}; therefore the observed trend could result from a difference in material density or porosity, assuming spherical particles. The advantage of this method of density determination is that both atmosphere and surface models utilize $\rho_{p}$ not $\rho_{m}$ and assuming our particles are spherical, we experimentally determine $\rho_{p}$. Further work is necessary to understand the porosity trends.

\section{Conclusions}
\begin{figure}
\plotone{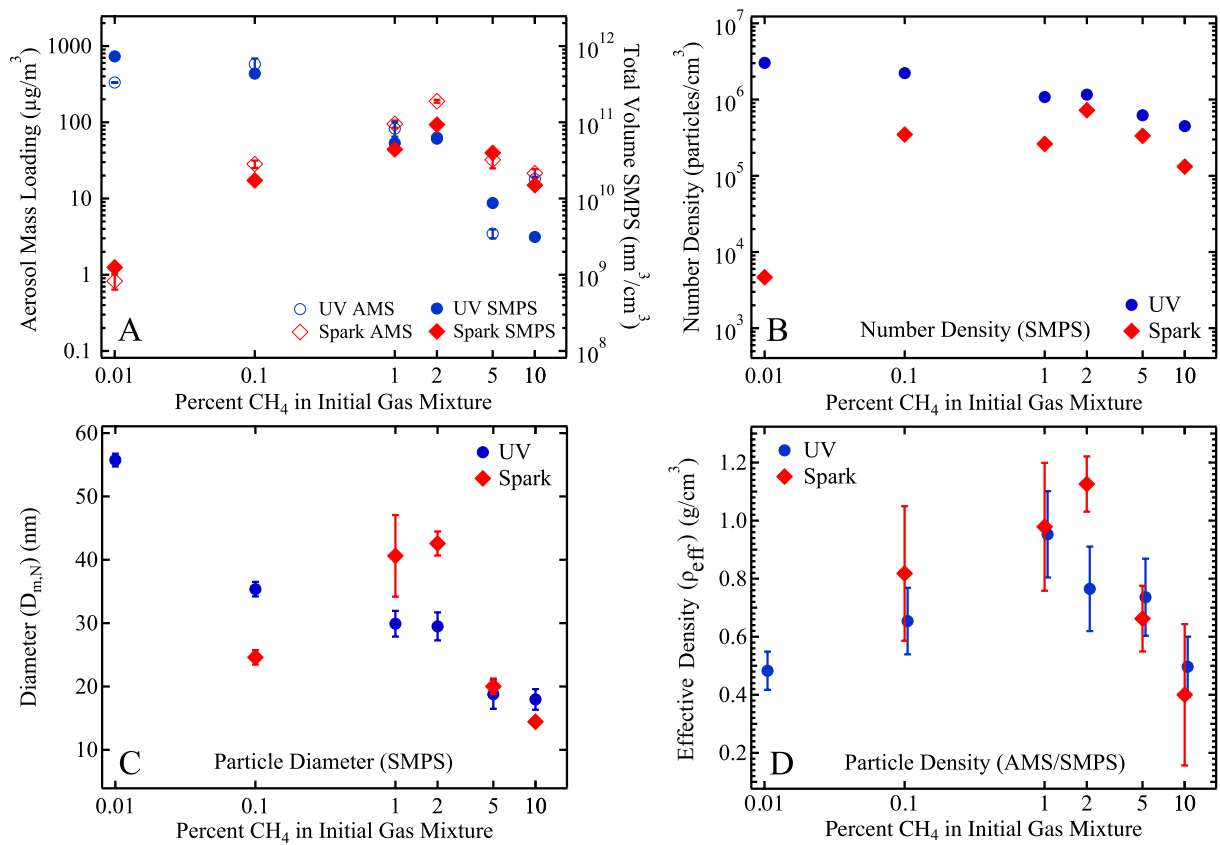}
\caption{AMS measurements of the total aerosol mass loading (left axis, solid symbols) and SMPS measurements of the total volume loading (right axis, open symbols) as a function of initial [CH$_{4}$] for the UV (blue, circles) and spark (red, diamonds) experiments are shown in Panel A. Number density and particle diameter measured by the SMPS are shown in Panels B and C, respectively. Effective particle density ($\rho_{eff}$) as calculated using Eq. 1 is shown in Panel D. Error bars represent 1$\sigma$ error on multiple measurements. \label{fig:allfour}}
\end{figure}

We obtained \emph{in situ} size, number density, and particle density measurements for tholins produced using [CH$_{4}$] from 0.01\% to 10\% and two different energy sources, spark discharge and UV, shown in Figure \ref{fig:allfour} and Table 1. The UV tholins show a decrease in particle size and number density with increasing [CH$_{4}$]. The UV tholin density peaks at 1\% CH$_{4}$ with a value of 0.95 $\pm$ 0.11 g/cm$^{3}$. The spark tholins have a peak in size, number density, and particle density at 2\% CH$_{4}$ with a density of 1.13 g/cm$^{3}$. For the most Titan relevant [CH$_{4}$], our spark and UV tholins are a factor of 2-3 times smaller than the particles observed in Titan's troposphere by DISR. Interestingly, with the exception of the 2\% CH$_{4}$ experiments, the densities at a given [CH$_{4}$] are the same for the spark or UV experiment within the error bars. 

With the exception of the 2\% CH$_{4}$ spark tholin, all of our tholins have densities lower than the standard value for particle density, 1 g/cm$^{3}$, used in Titan aerosol and cloud microphysics models. Additionally, all of our measured densities are significantly lower than 1.5 g/cm$^{3}$, a value typically used by surface models. The type and efficiency of fluvial transport depends on the density difference between the liquid and sediment \citep{Burr:2006}. Liquid CH$_{4}$ has a density of 0.45 g/cm$^{3}$ \citep{Lorenz:2003} and our densities would result in very different transport behavior; some of our tholins might even float. This work suggests that the values being used in standard Titan models could be too high. For a variety of reasons, including temperature, pressure, and energy source, tholins are not perfect analogues of the aerosols in Titan's atmosphere (see \citet{Cable:2012} for a detailed discussion), which is one of the plethora of reasons why we cannot rely on laboratory simulation experiments to provide final answers about Titan's aerosols. Rather we consider this work to be a first step in providing experimental constraints on parameters required by Titan models that are being used to interpret the results of Earth-based and spacecraft observations.

\acknowledgments
SMH is supported by NSF Astronomy and Astrophysics Postdoctoral Fellowship AST-1102827. This work was supported by NASA Planetary Atmospheres Grant NNX11AD82G.

\clearpage

\clearpage

\renewcommand{\baselinestretch}{1}

\begin{deluxetable}{llll}
\tabletypesize{\small} 
\tablenum{1} 
\tablecolumns{4}
\tablewidth{0pt}
\tablecaption{Summary of Particle Size, Number Density, and Particle Density\label{Table:data}}
\tablehead{\colhead{[CH$_{4}$] }&\colhead{Diameter}&\colhead{Number density}&\colhead{Effective density}\\
\colhead{(\%)}&\colhead{(D$_{m}$, nm)}&\colhead{(cm$^{-3}$)}&\colhead{($\rho_{eff}$, g/cm$^{3}$)}}
\startdata
Spark&&&\\
\hline
0.01$^{*}$&&4.6$\times10^{3}$ $\pm$ 2.2$\times10^{3}$&\\
0.1&24.6 $\pm$	1.1&3.5$\times10^{5}$ $\pm$ 6.4$\times10^{3}$& 0.82 $\pm$ 0.23\\
1&40.6 $\pm$ 6.5& 2.6$\times10^{5}$ $\pm$ 4.1$\times10^{3}$& 0.98 $\pm$ 0.22\\
2&42.6 $\pm$ 1.9& 7.2$\times10^{5}$ $\pm$ 1.6$\times10^{4}$& 1.13 $\pm$ 0.10\\
5&20.0 $\pm$ 1.2& 3.3$\times10^{5}$ $\pm$ 3.1$\times10^{3}$& 0.66 $\pm$ 0.11\\
10&14.4 $\pm$ 0.6& 1.3$\times10^{5}$ $\pm$ 9.5$\times10^{3}$& 0.40 $\pm$ 0.24\\
\hline
\hline
UV&&&\\
\hline
0.01&55.8 $\pm$ 1.0& 3.0$\times10^{6}$ $\pm$ 2.4$\times10^{3}$& 0.48 $\pm$ 0.07\\
0.1&35.4 $\pm$ 1.2& 2.2$\times10^{6}$ $\pm$ 1.9$\times10^{4}$& 0.65 $\pm$ 0.11\\
1& 29.9	$\pm$ 2.0& 1.1$\times10^{6}$ $\pm$ 2.0$\times10^{3}$&0.95 $\pm$ 0.15\\
2& 29.5 $\pm$	2.2 & 1.2$\times10^{6}$ $\pm$ 5.8$\times10^{3}$& 0.76 $\pm$ 0.15\\
5& 18.8 $\pm$	2.3 & 6.2$\times10^{5}$ $\pm$ 2.2$\times10^{2}$& 0.74 $\pm$ 0.13\\	
10& 18.0 $\pm$ 1.6 & 4.5$\times10^{5}$ $\pm$ 1.4$\times10^{4}$& 0.50 $\pm$ 0.10\\
\hline
\enddata
\tablecomments{$^{*}$The 0.01\% spark experiment produced very little aerosol. It was possible to obtain number density measurements, but the error on the size measurements was too large.}
\end{deluxetable}

\clearpage

{\renewcommand{\arraystretch}{0.3}
\setlength{\tabcolsep}{2pt}

\begin{deluxetable}{llll}
\tabletypesize{\small} 
\tablenum{2}
\tablecolumns{4}
\tablewidth{0pt}
\tablecaption{Summary of Relationships Between Effective Density ($\rho_{eff}$), Particle Density ($\rho_{p}$), and Material Density ($\rho_{m}$) as a Function of Particle Shape (adapted from \citet{Decarlo:2004})\label{table:density}}
\tablehead{\colhead{}&\colhead{}&\colhead{}&\colhead{}}
\startdata
&&&\\
&&&\\
&&\resizebox{0.6in}{!}{\includegraphics{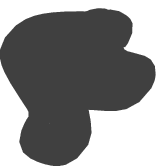}}&\raisebox{5.5ex}[0pt]{Irregular (no voids)}\\
&&&\raisebox{3.5ex}[0pt]{$\rho_{eff}<\rho_{p}=\rho_{m}$}\\
\raisebox{5.5ex}[0pt]{\resizebox{0.6in}{!}{\includegraphics{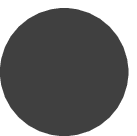}}}&\raisebox{11ex}[0pt]{Sphere (no voids)}&\resizebox{0.6in}{!}{\includegraphics{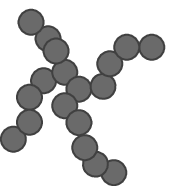}}&\raisebox{5.5ex}[0pt]{Aggregate (no voids)}\\
&\raisebox{9ex}[0pt]{$\rho_{eff}=\rho_{p}=\rho_{m}$}&&\raisebox{3.5ex}[0pt]{$\rho_{eff}<\rho_{p}=\rho_{m}$}\\
\raisebox{5.5ex}[0pt]{\resizebox{0.6in}{!}{\includegraphics{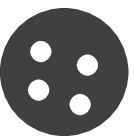}}}&\raisebox{11ex}[0pt]{Sphere (internal voids)}&\resizebox{0.6in}{!}{\includegraphics{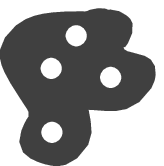}}&\raisebox{5.5ex}[0pt]{Irregular (internal voids)}\\
&\raisebox{9ex}[0pt]{$\rho_{eff}=\rho_{p}<\rho_{m}$}&&\raisebox{3.5ex}[0pt]{$\rho_{eff}<\rho_{p}<\rho_{m}$}\\
\raisebox{5.5ex}[0pt]{\resizebox{0.6in}{!}{\includegraphics{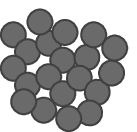}}}&\raisebox{11ex}[0pt]{Compact aggregate}&\resizebox{0.6in}{!}{\includegraphics{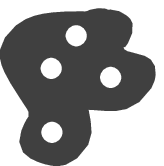}}&\raisebox{5.5ex}[0pt]{Aggregate (internal voids)}\\
&\raisebox{9ex}[0pt]{$\rho_{eff}\approx\rho_{p}<\rho_{m}$}&&\raisebox{3.5ex}[0pt]{$\rho_{eff}<\rho_{p}<\rho_{m}$}\\
\hline
\enddata
\end{deluxetable}

%\resizebox{0.6in}{!}{\includegraphics{particleg.eps}}
%% If the table is more than one page long, the width of the table can vary
%% from page to page when the default \tablewidth is used, as below.  The
%% individual table widths for each page will be written to the log file; a
%% maximum tablewidth for the table can be computed from these values.
%% The \tablewidth argument can then be reset and the file reprocessed, so
%% that the table is of uniform width throughout. Try getting the widths
%% from the log file and changing the \tablewidth parameter to see how
%% adjusting this value affects table formatting.

%% The \dataset{} macro has also been applied to a few of the objects to
%% show how many observations can be tagged in a table.

%% Tables may also be prepared as separate files. See the accompanying
%% sample file table.tex for an example of an external table file.
%% To include an external file in your main document, use the \input
%% command. Uncomment the line below to include table.tex in this
%% sample file. (Note that you will need to comment out the \documentclass,
%% \begin{document}, and \end{document} commands from table.tex if you want
%% to include it in this document.)


%% The following command ends your manuscript. LaTeX will ignore any text

\end{document}